# Cell transformation in tumor-development: a result of accumulation of Misrepairs of DNA through many generations of cells

Jicun Wang-Michelitsch[1]*, Thomas M Michelitsch[2]

[1]Department of Medicine, Addenbrooke's Hospital, University of Cambridge, UK (Work address until 2007)

[2] Institut Jean le Rond d'Alembert (Paris 6), CNRS UMR 7190 Paris, France

## Abstract

Development of a tumor is known to be a result of accumulation of DNA changes in somatic cells. However, the processes of how DNA changes are produced and how they accumulate in somatic cells are not clear. DNA changes include two types: point DNA mutations and chromosome changes. However, point DNA mutations (DNA mutations) are the main type of DNA changes that can remain and accumulate in cells. Severe DNA injuries are the causes for DNA mutations. However, Misrepair of DNA is an essential process for transforming a DNA injury into a "survivable and inheritable" DNA mutation. In somatic cells, Misrepair of DNA is the main source of DNA mutations. Since the surviving chance of a cell by Misrepair of DNA is low, accumulation of DNA mutations can take place only possibly in the cells that can proliferate. Tumors can only develop in the tissues that are regenerable. The accumulation of Misrepairs of DNA needs to proceed in many generations of cells, and cell transformation from a normal cell into a tumor cell is a slow and long process. However, once a cell is transformed especially when it is malignantly transformed, the deficiency of DNA repair and the rapid cell proliferation will accelerate the accumulation of DNA mutations. The process of accumulation of DNA mutations is actually the process of aging of a genome DNA. Repeated cell injuries and repeated cell regenerations are the two preconditions for tumor-development. For cancer prevention, a moderate and flexible living style is advised.

It is known that development of a tumor is a result of accumulation of DNA changes in somatic cells. A tumor cell is a transformed cell from a normal cell by DNA changes, which enable the cell to proliferate independently. Although we have had made enormous progress on diagnosis and therapy of some forms of tumors, the common mechanism of tumor development is not fully understood. The key processes in cell transformation, including how DNA changes are produced and how they accumulate in cells, are not clear. Most types of tumors are aging-associated; however traditional aging theories cannot interpret the phenomenon of tumor-development. In the present paper, we will demonstrate that our novel aging theory, the Misrepair-accumulation theory (Wang, 2009), is distinct in this aspect. The Misrepair mechanism that is proposed in this theory can explain the processes of production and accumulation of DNA mutations in cell transformation. Our discussion tackles the following issues:

I. Characteristics of tumor cells and tumor-development

    1.1 Independence of tumor cells on dividing and surviving
    1.2 Tissue-selectivity in tumor-development

II. A generalized concept of Misrepair

III. Production and accumulation of DNA mutations in somatic cells

    3.1 Misrepair of DNA: the main source of DNA mutations in somatic cells
    3.2 Accumulation of DNA mutations: through many generations of cells
    3.3 Accumulation of DNA mutations in malignant tumor cells: self-accelerating and inhomogeneous
    3.4 Aging of a genome DNA

IV. Successive acquisitions of new properties of tumor cells

V. Preconditions for cancer-development: repeated cell injuries + repeated cell-regenerations

VI. Conclusions

## I. Characteristics of tumor cells and tumor-development

Tumor-development is a result of uncontrollable cell proliferation of an immortal cell. An immortal cell is a transformed cell from a normal somatic cell after acquiring new properties through DNA mutations. Rapid expansion of a tumor tissue, invasion of tumor cells into neighbor tissues, and metastasis of tumor cells to other locations through blood circulation ..., all these behaviors of a tumor may lead to a rapid failure of one or multiple organs. To understand the common mechanism of cell transformation, it is essential to analyze the behaviors of tumor cells and the manners of tumor-development. Apart from the properties of tumor cells, tumor-development has two characteristics: age-related and tissue-selected. In our view, tissue-selectivity of tumors gives us an important clue for studying the mechanism of cell transformation.

## 1.1 Independence of tumor cells on dividing and surviving

Except blood cells, all the cells in an organism need to survive in a tissue environment where they can communicate with each other directly or via extracellular matrixes (ECMs). The cells in a tissue are interdependent on each other for functioning and for surviving. Loss of contact with neighbor cells/ECMs will lead to cell death, and this is called "loss of anchor apoptosis" or "anoikis". A normal stem cell needs to be stimulated by external signals for proliferation, and cell division will be terminated by withdrawing of the stimulating signals and/or by a stopping signal. Cell-contact is a universal stopping signal for terminating cell proliferation, and this phenomenon is called "cell-contact inhibition". The behaviors of normal cells in a tissue are strictly and precisely controlled by their tissue environment, and none of them is independent.

Differently, a tumor cell has lost its dependence on other cells for proliferation and even for survival. It can undergo cell division without being stimulated, and the cell division cannot be stopped by cell-contact. With these two properties, namely, the property of stimulator-independent mitotic division and the property of loss of cell-contact inhibition, a tumor cell can proliferate unlimitedly into a clone, forming a tumor. Apart from these two properties, the tumor cells in a malignant tumor have additional properties, including production of matrix metalloproteinase (MMP), anoikis-resistance, and acquired mobility. With the ability of producing MMPs to digest ECMs, the tumor cells can invade into neighbor tissues. With the property of anoikis-resistance, a tumor cell can survive without anchoring to neighbor cells/ECMs, and it can immigrate passively to other organs via blood circulation. A tumor cell can migrate actively if it has additionally obtained mobility. Therefore, these five properties are the most important properties for tumor cells:

- Stimulator-independent mitotic division
- Loss of cell-contact inhibition
- MMP-production
- Anoikis-resistance
- Acquired mobility

## 1.2 Tissue-selectivity in tumor-development

Tumors are classified into two groups with respect to the ages of onset of tumors: aging-associated tumors and genetic tumors. Aging-associated tumors are the tumors that develop mainly in the people age over 50 years old, with increasing incidence with age. Most of human tumors are aging-associated, including lung cancer, gastric cancer and colorectal cancer. Genetic tumors are the tumors that develop mainly in young children but rarely in adults. Genetic tumors are often rare tumors, such as neuroblastoma, neurofibromatosis, and muscle tissue sarcoma. Genetic tumors develop often in the tissues where aging-associated tumors do not develop. In contrast, aging-associated tumors mainly develop from epithelial cells (including glandular cells), hepatocytes, and glial cells, from which genetic tumors hardly develop. The phenomenon of tissue-selectivity in tumor-development suggests that aging-associated tumors develop mainly in the tissues that are regenerable. Normally, a DNA change can occur to any one of somatic cells, but why does cell transformation take place only in regenerable cells?

Tumor-development is a part of aging of an organism. A reasonable aging theory should be able to interpret this phenomenon. However, traditional biological theories including gene-control theory (Fabrizio, 2010; McCormick, 2012) and damage (fault)-accumulation theory (Kirkhood, 2005) fail to explain tumor-development. Tumors have a great diversity on cell-origin, and different individuals may develop the same form of tumor at different ages. Therefore, development of a tumor may not be possibly a result of controlling of certain genes, which should work in the same way in different individuals. Damage (fault)-accumulation theory emphasizes that it is the "intrinsic faults", due to the limitation of repair/maintenance, that accumulate and lead to aging (Kirkhood, 2005). However, it is found that the DNA mutations in somatic cells are mainly produced by Mis-match repair (Misrepair) of DNA rather than by "faults" (Natarajan, 1993; Bishay, 2001). For a cell that suffers from a DNA injury, lack of DNA repair would lead to cell death rather than DNA mutation. On interpreting the phenomenon of tumor-development, these two leading aging theories are not tenable. Differently, with a concept of Misrepair mechanism, the Misrepair-accumulation theory is able to explain tumor-development by showing how DNA mutations are produced and how they accumulate in somatic cells.

## II.  A generalized concept of Misrepair

The term of Misrepair is traditionally used on DNA, referred to the Mis-match repair of DNA. Misrepair of DNA is a strategy of DNA repair when a DNA is severely injured. Actually, incorrect repairs take place also on cells and tissues, and scar formation is an example. On this basis, we proposed in our Misrepair-accumulation theory a generalized concept of Misrepair for describing all types of incorrect repairs. The new concept of Misrepair is defined as *incorrect reconstruction of an injured living structure,* and it is applicable to all living structures including molecules (DNA), cells, tissues (Wang, 2009). DNA Misrepair is a special case of Misrepair. In situation of a severe injury, when complete repair is impossible to achieve, Misrepair, a repair with altered materials and in altered remodeling of a structure, is a way to maintain the structural integrity and increase the surviving chance of an organism. Without Misrepairs, an individual could not possibly survive to the age of reproduction; thus Misrepair mechanism is a surviving mechanism for an organism and for a species.

Misrepairs have three characteristics: unavoidable, with alternation of structure, and irreversible. Therefore Misrepairs will accumulate in an organism. Accumulation of Misrepairs will gradually disorganize the structure of a molecule, a cell, or a tissue, appearing as aging of it. Misrepairs have a tendency to accumulate focally in a tissue, because an old Misrepair makes local part of the tissue have reduced repair-efficiency and increased damage-sensitivity. Accumulation of Misrepairs is thus self-accelerating, focalized, and inhomogeneous. Aging may take place on the levels of molecules, cells, and tissues respectively; however aging of a multi-cellular organism takes place essentially on tissue level. Although remaining of aged cells can be part of the aging of a tissue, an irreversible change of the spatial relationship between cells/ECMs in a tissue is *essential and sufficient* for leading to a decline of tissue functionality and body functionality. Aging of an organism does not always require aging of cells. In summary, aging of an organism is a result of accumulation of Misrepairs on tissue level. Misrepair mechanism is beneficial for the survival of a species, and this is the evolutionary advantage of aging mechanism. Aging of an individual is a sacrifice for the species' survival.

### III. Production and accumulation of DNA mutations in somatic cells

DNA changes are the genetic basis for tumor-development and they are responsible for the alteration of the phenotypes of tumor cells. The type and the amount of DNA changes are different in different forms of tumors and in different patients. It would be too time-consuming and expensive to study all the DNA mutations in each patient. More important is to study the common mechanism in all tumors on: **A.** how DNA changes are produced in somatic cells and **B**. how DNA changes accumulate in cells.

### 3.1  Misrepair of DNA: the main source of DNA mutations in somatic cells

Point DNA mutations and chromosomal changes are the two types of DNA changes that can occur in a cell. Chromosomal changes on number and/or on structure are often fatal to tissue cells; thus they cannot remain and accumulate in cells. Differently, a point DNA mutation (called DNA mutation or gene mutation) on one or two bases of a DNA is often silent or mild, and it can "survive" and accumulate in cells. Therefore, DNA mutations are the main DNA changes that can accumulate for long time and contribute to the cell transformation in older age. On the cause for generation of DNA mutations, there is still a debate. For some biologists, a DNA mutation is a result of intrinsic error of DNA synthesis. However, this idea is untenable. If a DNA mutation is produced by an error during DNA duplication, accumulation of DNA mutations would be much rapider in a growing child than in an adult. If it is like this, tumor-development would have much higher incidence in children than in adults, which is not true. For some other scientists, DNA mutations are a kind of DNA "damage". This idea is also misleading. An original DNA damage (injury) is in fact a DNA break, and a cell cannot survive if a broken DNA is not re-linked. Thus, DNA "damage" cannot remain and accumulate in a living cell. A broken DNA must be re-linked for the survival of a cell, even if sometimes it can be re-linked incorrectly.

A DNA injury (break) is the promoter of generation of a DNA mutation in a somatic cell (Kasparek, 2011). DNA break can be made by radiation, DNA-damaging chemicals, or viral attack. A DNA molecule is composed of double strands that are complementary on the base-sequence and attach to each other by hydrogen bonds. One strand can be used as the template for producing the other strand. This is the structural basis for DNA synthesis and DNA repair.

When a DNA break is small, it can be fully repaired. For example, when one DNA strand is broken, the broken strand can be re-linked by new bases in correct sequence by using the uninjured strand as the template. However, when double strands of a DNA disrupt in the same time, no template can be used for re-linking the DNA correctly. In this case, other repair pathways such as non-homologous end joining of broken DNAs have to be promoted for re-linking the DNA (Moore, 1996; Kasparek, 2011). Otherwise the cell will die from failure of DNA. Such a repair may result in alteration of local DNA sequence, such as alternation, deletion, or insertion of one or two bases. Therefore the repair in this way is a "Misrepair". In the following discussion, the term of "Misrepair of DNA" is referred to the process of Misrepair of DNA, and the term of "DNA Misrepair" is referred to the result of that, namely, a DNA mutation.

Misrepair of DNA is an essential process for transforming a DNA break into a "survivable and inheritable" DNA mutation. Misrepair of DNA is a strategy of DNA repair but not a result of failure of repair. Misrepair of DNA can only take place when a cell has normal function of DNA repair. In contrast, for a cell with defect of DNA repair, a DNA injury will lead to cell death rather than a DNA mutation. Like the Misrepairs in other aging changes, DNA Misrepairs are made for increasing the surviving chance of a cell; however they may exhibit their side-effects later in tumor-development. Studies have shown that Misrepair of DNA is the main source of DNA mutations. For example, the misrepaired double-strand breaks are found to be the main lesions as the origin of both of chromosomal abnormalities and gene mutations (Natarajan, 1993; Bishay, 2001). Misrepair of DNA is one of the surviving mechanisms of a cell under radiation, but is also the origin of tumor-development (Rothkamm, 2002). The "misrepaired DNA damage" is a causal factor for gene mutation and development of cancer and aging (Suh, 2006).

Although Misrepair of DNA is a strategy for cell survival, the surviving chance of a cell from Misrepair of DNA is low. There are three reasons for that: **A.** a cell that has DNA injuries can have also severe injuries on other parts of the cell; **B.** some DNA Misrepairs are fatal to cells; and **C.** the cells with altered phenotypes by DNA Misrepair will be removed by immune system. Most of the cells with severe DNA injuries will die, and only quite a few may survive by Misrepair. Therefore, production of a DNA mutation is a rare affair, and this is often paid by death of a large number of cells.

### 3.2 Accumulation of DNA mutations: through many generations of somatic cells

Since the chance of a cell to survive from Misrepair of DNA is low, the opportunity of one cell to survive two times of Misrepair of DNA is almost zero. Accumulation of DNA Misrepairs in one cell is hardly possible. However, for the cells that are reproducible, namely the stem cells, DNA mutations can be inherited and accumulate in their offspring cells. Therefore, accumulation of DNA mutations can only proceed possibly in reproducible cells, and this is one reason why tumors develop selectively in regenerable tissues. For example, for part of a tissue that is frequently exposed to certain damage, if the surviving chance of a cell from this damage by Misrepair of DNA is 1‰, the surviving chance of one cell from two times of DNA injury is almost zero, namely 1‰ x 1‰ = 1/1000000. However, if a stem cell, which has survived by a DNA Misrepair (the first Misrepair), has reproduced $2 \times 10^3 \times 10^3$ (= $2 \times 1000000$) offspring cells, all of these cells will inherit this DNA Misrepair (mutation). When these cells suffer again from the same kind of DNA injury, some of them, possibly one or two cells, may survive by another Misrepair of DNA (the second Misrepair). Thus, DNA Misrepairs accumulate in these survived cells. To have this number of offspring cells, namely $2 \times 10^6$, a stem cell needs to proliferate into at least 17 generations ($2 \times 10^6 = 2 \times 2^{16} = 2^{17}$). By cell proliferation, more offspring cells will inherit these two DNA mutations, and some of them may possibly survive by a third DNA Misrepair. In this way, by repeated DNA injuries and repeated cell proliferations, more and more DNA mutations are produced and accumulate in the offspring cells from one mother cell. Finally one of the cells can be transformed after obtaining a new property (Figure 1). Thus, accumulation of DNA mutations needs to proceed over many generations of cells. For example, for accumulating ten times of DNA mutations by Misrepair of DNA, a cell needs to proliferate into at least 86 generations ($2 \times 10^{3 \times 10} = 2 \times 10^{6 \times 5} = 2 \times 2^{16 \times 5} = 2 \times 2^{85} = 2^{86}$).

Cell proliferation in a normal tissue is accurately controlled; and it needs to be promoted by stimulators for adapting to different situations. Thus, a stem cell and its offspring cells, which we call "a group of cells", need many years for proliferating into sufficient generations. If there are more numbers of stem cells in a local tissue, the load of DNA damage will be shared by "more groups of cells" and the risk of accumulation of DNA Misrepairs in "one group of cells" will be reduced. Therefore, in somatic cells, accumulation of DNA mutations is a slow process, and cell transformation needs a long time. As human being, we seem to have higher incidence of tumors than animals. One reason is that we live longer than them, from which we have longer time for accumulation of DNA mutations! All of us will develop a cancer if we live sufficiently long. Simpler organisms such as insects and worms do not develop cancers, because they live too short time (Campisi, 2000)! Low-dose aspirin is found to be effective in reducing the risk of tumor-development (Brotons, 2014). One mechanism may be: cell repair including repair and Misrepair of DNA is partially inhibited by aspirin, by which the process of accumulation of DNA Misrepairs is slowed down in the cells.

An accelerating element for accumulation of DNA mutations is the repeated exposure for a long term to radiation or DNA-toxic substances. Rapid accumulation of DNA mutations and high incidence of cancers may take place to the people, who have been exposed to intensive nuclear radiation like that in Chernobyl Disaster. Slow accumulation of DNA mutations is a result of daily exposure for many years to similar aggressive substances in food or in air. This is one reason why gastric cancer, colorectal cancer and lung cancer have high incidence in old people. Differently, genetic tumors develop mainly in the children who have had inherited gene mutations from their parents. A genetic DNA mutation exists in every somatic cell, and the defect on cell functionality caused by genetic mutations can make the accumulation of DNA mutations start earlier and proceed more rapidly. Cell transformation can therefore take place earlier in these individuals.

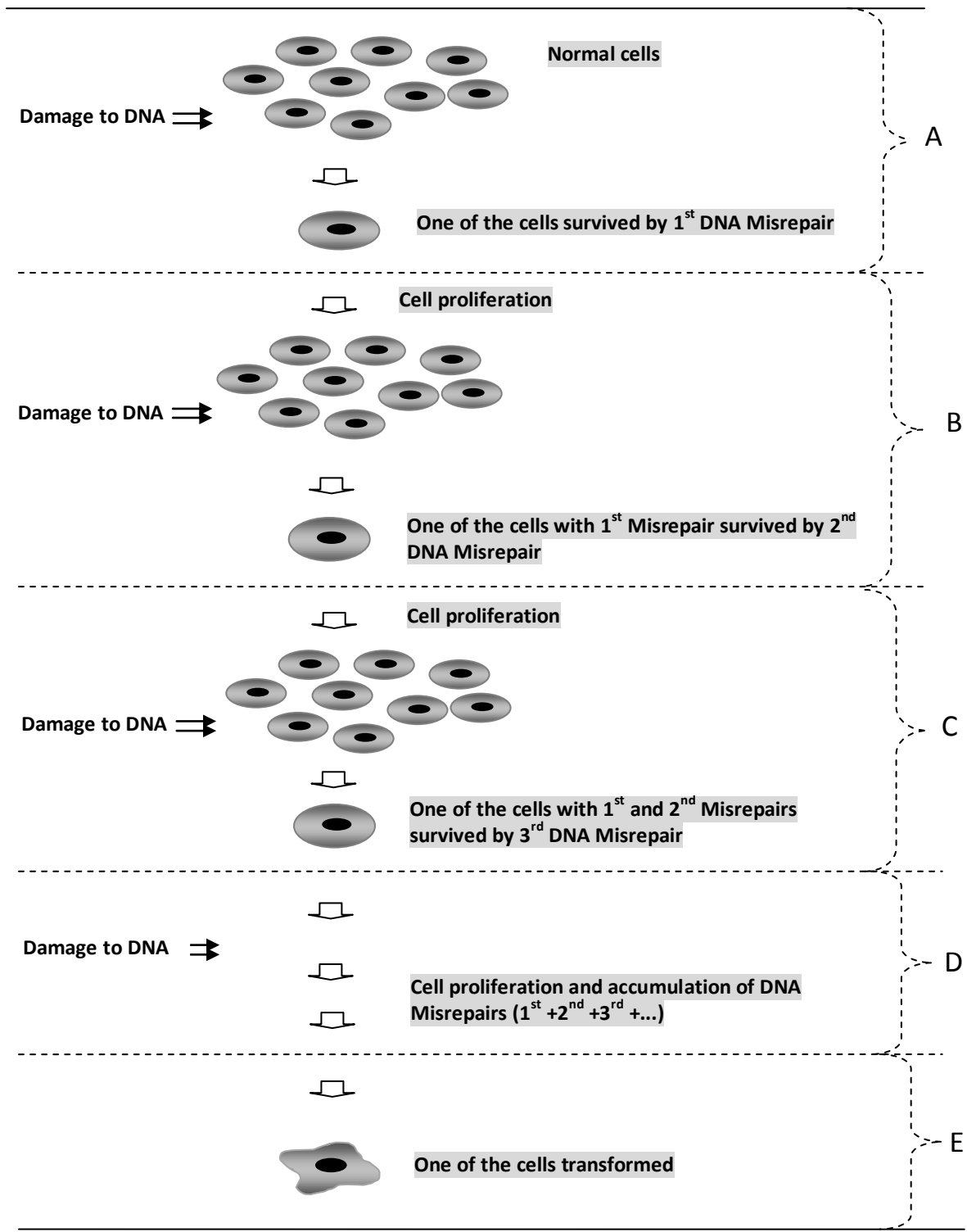

**Figure 1. Accumulation of DNA Misrepairs (mutations): proceeding over many generations of cells**

Most of the cells that suffer from severe DNA injuries will die, and only a few may survive through Misrepair of DNA (**A**). If a cell can proliferate, the DNA Misrepair (**the 1st Misrepair**) in the cell can be inherited by its offspring cells. If some of the offspring cells can survive from a DNA injury by another DNA Misrepair (**the 2nd Misrepair**), DNA Misrepairs can accumulate in these cells (**B**). By cell proliferation, more offspring cells will inherit these two DNA mutations, and some of them may possibly survive by **a 3rd DNA Misrepair**. In this way, by repeated cell proliferation and repeated DNA injuries, DNA mutations are produced and accumulate in the offspring cells of one mother cell (**D**). Finally one of the cells can be transformed after obtaining a new property (**E**). Accumulation of DNA mutations need to proceed over many generations of cells. Cell transformation of a normal cell to a tumor cell is a long and slow process.

## 3.3 Accumulation of DNA mutations in malignant tumor cells: self-accelerating and in-homogenous

Malignant tumor cells are different from normal cells on several aspects. **I.** Normal cells have full differentiation with full efficiency on DNA repair and other functions; whereas tumor cells have low degree of cell differentiation with functional deficiency on cell repair and on DNA repair. Because of deficiency of DNA repair, SOS repair pathway for re-linking DNA such as NHEJ may have to be often promoted in malignant tumor cells. To some DNA injuries, full repair can be achieved in a normal cell, but may not in a malignant tumor cell. The frequency of Misrepairs in malignant tumor cells is higher than that in normal cells. **II.** Cell proliferation of a normal stem cell is strictly controlled, whereas that of tumor cells is out of control and unlimited. **III.** The tolerance of a normal cell to a dominant DNA mutation is low, since an abnormal phenotype will lead to cell death or apoptosis. Differently, a malignant tumor cell can be more tolerant to DNA mutations because of immaturity of functionality. The surviving chance of a tumor cell from a DNA injury though Misrepair of DNA is higher than that of a normal cell. For example, a malignant tumor cell being anoikis-resistant is able to survive even if it cannot produce some proteins that are essential for cell-anchoring (Table 1). Taken together, accumulation of DNA mutations is accelerated in malignant tumor cells by rapider cell proliferation, reduced efficiency of DNA repair, and higher tolerance to mutations. The malignancy of a tumor cell is a result of DNA mutations; and in return, the malignancy will increase the frequency of DNA mutations in this group of cells. Accumulation of DNA mutations is therefore self-accelerating in the same group of tumor cells. Most of the mutations that are detected in malignant tumor cells could be produced during the progression of the tumor.

**Table 1. Accumulation of DNA mutations: rapider in malignant tumor cells than in normal cells**

|  | **Normal cells** | **Malignant tumor cells** |
| --- | --- | --- |
| Cell differentiation | High | Low |
| DNA repair | Full | Deficient |
| Cell proliferation | Well-controlled | Unlimited |
| Tolerance to DNA mutations | Low | High |

In a malignant tumor, accumulation of DNA mutations is more and more rapid, but the rates of accumulation of DNA mutations can be different in different tumor cells. During the growing of a tumor, tumor cells differentiate when some of them obtain new mutations, and sub-groups or sub-sub-groups of tumor cells with different mutations appear. As shown in Figure 2, although all of the tumor cells in a tumor have some common DNA mutations, such as mutation A (MA), they can be different on other mutations. For example, some cells with MA have additional mutations respectively: MB1, MB2 or MB3, and some cells with (MA+MB1) have additional mutations respectively: MC1, MC2 or MC3. The difference on the type and on the number of DNA mutations makes the cells into different sub-groups, such as sub-groups of (MA+MB1), of (MA+MB2), and of (MA+MB3), and even sub-sub-groups, such as sub-sub-groups of (MA+MB1+MC1), of (MA+MB1+MC2), and of (MA+MB1+MC3). Therefore, the accumulation of DNA mutations is in-homogenous in different sub-groups of tumor cells.

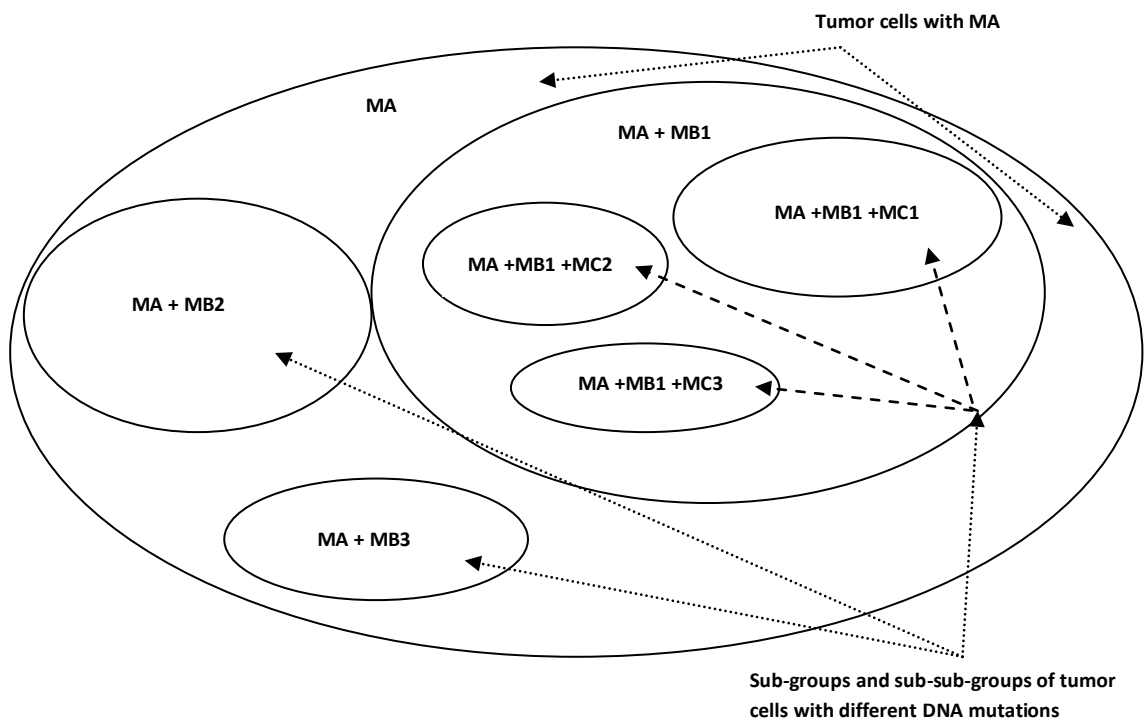

**Figure 2. In-homogenous accumulation of DNA mutations in different sub-groups of tumor cells**

In a malignant tumor, the rates of accumulation of DNA mutations can be different in different tumor cells. With the growing of a tumor, tumor cells will differentiate when some of them obtain new mutations, and sub-groups and sub-sub-groups of tumor cells with different mutations appear. For example, some cells with **MA** have additional mutations respectively: **MB1, MB2** and **MB3**, and some cells with (**MA+MB1**) have additional mutations respectively: **MC1, MC2** and **MC3**. The difference on type and on number of DNA mutations makes the cells into different sub-groups, such as sub-groups of (**MA+MB1**), of (**MA+MB2**), and of (**MA+MB3**), and even sub-sub-groups, such as sub-sub-groups of (**MA+MB1+MC1**), of (**MA+MB1+MC2**), and of (**MA+MB1+MC3**). Therefore, the accumulation of DNA mutations is in-homogenous in different sub-groups of tumor cells.

## 3.4  Aging of a genome DNA

Accumulation of DNA mutations can be understood as a process of aging of a genome DNA, since this process alters gradually the structure as well as the functionality of the genome DNA. Similarly to a scar on the skin, a DNA Misrepair (mutation) is a kind of aging change of DNA. However, aging of a DNA needs to proceed over many generations of (daughter) DNAs in many generations of cells. In somatic cells, aging of a genome DNA is slow, but in malignant tumor cells, that is rapid. Aging of a DNA may lead to tumor-development, and death of the organism from a cancer will make all aged genome DNAs disappear.

## IV. Successive acquisitions of new properties for cell transformation

The new cell properties in transforming a somatic cell into a tumor cell are essentially acquired by changes of DNA in the cell. The most important cell properties in cell transformation are stimulator-independent mitotic division, called here as property A (**PA**), loss of cell-contact inhibition (**PB**), MMP-production (**PC**), anoikis-resistance (**PD**), and acquired mobility (**PE**). Each of these properties can be a final effect of alterations of multiple genes by DNA mutations. Acquiring of a new cell property by DNA mutations is a complex process, and it may be different in different tumors and in different individuals. However, important is to know whether these properties are acquired successively in a certain sequence or simply in a random way. Theoretically, the occurrence of a DNA mutation and gain of a new property of a cell are random. However, for tumor-development, new mutations and new properties should be able to remain and accumulate in cells. A mutation in a single cell will disappear, when this cell dies before cell division. Thus, a mutation will have increased chance to remain if it takes place in the cells that can proliferate. An ideal sequence of acquiring of new properties for tumor-development should be the sequence that is beneficial for accumulation of new mutations. Based on this logic, we find out that one of the best sequences is: **PA** to (**PA+PB**), then to (**PA+PB+PC**), then to (**PA+PB+PC+PD**), and finally to (**PA+PB+PC+PD+PE**) (Figure 3).

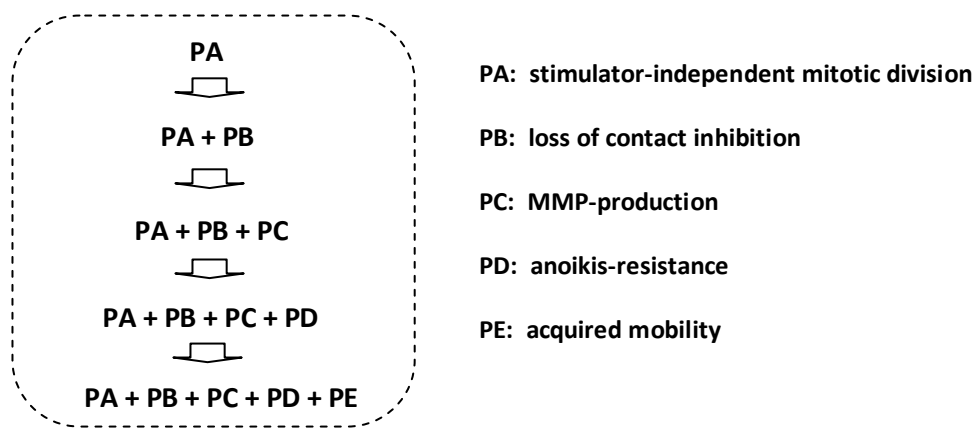

**Figure 3. Hypothesized sequence of acquiring of new properties of a cell for transformation**

The most important properties of tumor cells are stimulator-independent mitotic division, the property A (**PA**), loss of cell-contact inhibition (**PB**), MMP-production (**PC**), anoikis-resistance (**PD**), and acquired mobility (**PE**). These properties may be acquired by a normal cell successively in a certain sequence. In our view, one of the best sequences of acquiring of these properties for cell transformation is: **PA** to (**PA+PB**), then to (**PA+PB+PC**), then to (**PA+PB+PC+PD**), and finally to (**PA+PB+PC+PD+PE**).

The first property to be acquired should be **PA**, since this property helps make copies of a DNA mutation in daughter cells. In the cells with this property, the DNA mutations for **PA** and other mutations that take place later will have increased chance to remain. Although cell proliferation is still restricted by cell-contact inhibition in this stage, the cells can proliferate in certain environments where there is not this inhibition. In contrast, if a cell had at first acquired one of other properties, the corresponding DNA mutations cannot remain when the single carrying cell dies. Without **PA**, accumulation of new properties is difficult in cells. The second property should be **PB**, since this additional property makes the cells with **PA** overcome the limit of cell-contact inhibition for proliferation. The properties of **PA+PB** will increase the cell number and the copies of DNA mutations, and can transform a normal cell into a tumor cell. Other three properties, including **PC**, **PD**, and **PE**, are not essential for all tumors, but essential for malignant tumors. **PC** and **PD** should be acquired earlier than **PE**, because **PC** and **PD** are sufficient for the aggressive behavior of a tumor. **PC,** by digesting ECMs in tissues, enables the local invasion of a tumor. **PD** enables a cell survive even if it has lost its contact with neighbor cells. Therefore the two properties, **PC** and **PD,** are both needed for a faraway metastasis of a tumor cell. Taken together, **PA** and **PB** are the two essential properties for all types of tumor cells, and they should be obtained before clonal evolution. **PC, PD**, and **PE** are possibly acquired successively during the procession of a tumor from pre-malignant to malignant.

## V. Preconditions for cancer-development: repeated cell injuries + repeated cell regenerations

Repeated cell injuries are the triggers for DNA mutations, and cell proliferation enables the accumulation of DNA mutations in cells. Thus, repeated cell injuries and repeated cell regenerations are the two preconditions for tumor-development. For example, gastric cancer and colorectal cancer are results of repeated injuries of mucosa and repeated regenerations of epithelial cells for repair. Development of breast cancer may be a consequence of repeated proliferation-regression of mammary glands. Proliferation of mammary cells is promoted by increase of hormones; and death of lobule cells is promoted by decrease of these hormones. DNA mutations may take place when some "strong" cells survive from "the edge of death". The DNA mutations in the stem cells of terminal ducts of breast glands may accumulate after many circles of proliferation-regression. Some forms of cancers such as colorectal cancer and breast cancer have in some cases genetic predisposition. One of the genetic impacts on oncogenesis is by enhancing cell proliferation. High level of hormones, which can promote cell proliferation, is found to be a cause for the high sensitivity of some people to cancers. For example, the high risk of breast cancer in the daughters of the women who have had breast cancer might be due to the high levels of estrogen and progesterone in these individuals (Henderson, 2000).

For cancer prevention, it is therefore important to reduce the opportunity of repeated injuries to the same part of an organ. We need to make efforts from several aspects. **Firstly**, it is better live in a moderate style for avoiding unnecessary damage-exposure such as smoking, too much eating, prolonged sun-bath, and violent sports. **Secondly**, we should therapy actively acute inflammations for preventing them from becoming chronic. **Thirdly**, we need to reduce the opportunity of a long-term exposure to the same type of damage. Some substances in food or in air are probably not damaging directly, but the half-degraded products of them can be aggressive. It is reported that gastric cancer has high incidence in Asia whereas colorectal cancer has high incidence in Europe. The difference on the habits on food-consummation between Asians and Europeans can be a reason for this difference on tumor-development (Brenner, 2009). Therefore, increasing the diversity of foods can be a way to reduce the risk of these two forms of cancers. When we have a large diversity of foods, the opportunity of damage of certain food to the same part of gastric-intestinal mucosa will be reduced. That is to say, the load of damage will be shared by more groups of cells in different parts of digestive mucosa if we have more kinds of foods. Thus, when we have a flexible life on eating, on moving, and on working, we will have reduced risk to be exposed to the same type of damage, physical or chemical. Taken together, a moderate and flexible living style can be helpful in reducing the risk of cancer-development.

## VI. Conclusions

In somatic cells, DNA mutations are mainly produced by Misrepair of DNA. Accumulation of DNA mutations can only proceed in the cells that are reproducible. This is one reason why tumor-development takes place mainly in the tissues that are regenerable. Accumulation of DNA mutations needs to proceed over many generations of cells; therefore cell transformation from a normal cell into a tumor cell is a slow and long process. However, once a cell is transformed, especially when it is malignantly transformed, rapid cell proliferation and deficiency of DNA repair will accelerate the accumulation of DNA mutations. Accumulation of DNA mutations is in fact the process of aging of a genome DNA. Cell transformation can be understood as a process of cell evolution. This process is similar to that in species' evolution. In both of cell evolution and species' evolution, the mutations (variations) are random, tiny, but accumulating due to their inheritability. Repeated cell injuries and repeated cell regenerations are the two preconditions for tumor-development. For cancer prevention, a moderate and flexible living style is beneficial.

**References**

1. Bertola, Dr; Cao, H; Albano, Lm; Oliveira, Dp; Kok, F; Marques-Dias, Mj; Kim, Ca; Hegele, Ra (2006). "Cockayne syndrome type A: novel mutations in eight typical patients". Journal of human genetics 51 (8): 701–5. doi:10.1007/s10038-006-0011-7. PMID 16865293.
2. Bishay K, Ory K, Olivier MF, Lebeau J, Levalois C, Chevillard S. (2001) DNA damage-related RNA expression to assess individual sensitivity to ionizing radiation. Carcinogenesis. 22(8), 1179
3. Brenner H, Rothenbacher D, Arndt V (2009). Epidemiology of gastric cancer. Methods Mol. Biol. Methods in Molecular Biology 472: 467–77. doi:10.1007/978-1-60327-492-0_23. ISBN 978-1-60327-491-3. PMID 19107449.

4. Brotons C, Benamouzig R, Filipiak KJ, Limmroth V, Borghi C. A Systematic Review of Aspirin in Primary Prevention: Is It Time for a New Approach? Am J Cardiovasc Drugs. 2014 Dec 12
5. Croce CM. (2008). "Oncogenes and cancer". N Engl J Med. 358 (5): 502–11.
6. Fabrizio P, Hoon S, Shamalnasab M, Galbani A, Wei M, Giaever G, Nislow C, Longo VD. (2010). Genome-wide screen in Saccharomyces cerevisiae identifies vacuolar protein sorting, autophagy, biosynthetic, and tRNA methylation genes involved in life span regulation. *PLoS genetics* **6**(7):e1001024.
7. Hanahan, Douglas; Weinberg, Robert A. (2000). The hallmarks of cancer. Cell 100 (1): 57–70. doi:10.1016/S0092-8674(00)81683-9.
8. Henderson BE, Bernstein L, Ross RK. (2000). Chapter 13: Hormones and the Etiology of Cancer ". In Bast RC, Kufe DW, Pollock RE, et al. Holland-Frei Cancer Medicine (5th ed.). Hamilton, Ontario: B.C. Decker. ISBN 1-55009-113-1. Retrieved 27 January 2011.
9. Kasparek TR, Humphrey TC. DNA double-strand break repair pathways, chromosomal rearrangements and cancer. Semin Cell Dev Biol. 2011 Oct; 22(8):886-97. doi: 10.1016/j.semcdb.2011.10.007. Review. PMID: 22027614
10. Kirkwood, T. (2005). Understanding the odd science of aging. Cell, 120.
11. McCormick M, Chen K, Ramaswamy P, and Kenyon C. (2012). New genes that extend Caenorhabditis elegans' lifespan in response to reproductive signals. *Aging Cell* **11**(2):192-202
12. Moore JK, Haber JE. (1996) Cell cycle and genetic requirements of two pathways of nonhomologous end-joining repair of double-strand breaks in Saccharomyces cerevisiae. Mol Cell Biol.16(5):2164-73. PMID : 8628283
13. Natarajan,A.T. (1993). Mechanisms for induction of mutations and chromosome alterations. Environ. Health Perspect., 101 (suppl. 3), 225
14. Rothkamm K, Löbrich M. Misrepair of radiation-induced DNA double-strand breaks and its relevance for tumorigenesis and cancer treatment (review). Int J Oncol. 2002 Aug;21(2):433-40. Review. PMID:12118342
15. Suh Y, Vijg J. (2006) Maintaining genetic integrity in aging: a zero sum game. Antioxid Redox Signal. 8(3-4), 559
16. Wang J, Michelitsch T, Wunderlin A, Mahadeva R. (2009) Aging as a Consequence of Misrepair – a Novel Theory of Aging. ArXiv: 0904.0575. arxiv.org